# A Hybrid Quantum–Classical Pipeline for X-Ray-Based Fracture Diagnosis

[1]Sahil Tomar, [2]Rajeshwar Tripathi, [3]Sandeep Kumar
[1,2,3] Central Research Laboratory, BEL, Ghaziabad, India

*Abstract*— Bone fractures are a leading cause of morbidity and disability worldwide, imposing significant clinical and economic burdens on healthcare systems. Traditional X-ray interpretation is time-consuming and error-prone, while existing machine learning and deep learning solutions often demand extensive feature engineering, large, annotated datasets, and high computational resources. To address these challenges, a distributed hybrid quantum–classical pipeline is proposed that first applies Principal Component Analysis (PCA) for dimensionality reduction and then leverages a 4-qubit quantum amplitude-encoding circuit for feature enrichment. By fusing eight PCA-derived features with eight quantum-enhanced features into a 16-dimensional vector and then classifying with different machine learning models achieving 99% accuracy using a public multi-region X-ray dataset on par with state-of-the-art transfer learning models— while reducing feature extraction time by 82%.

*Index Terms*— Bone fractures Detection, Principal Component Analysis, Quantum Feature extraction, X-Ray Medical Imaging.

## I. INTRODUCTION

Bone fractures present a major challenge in orthopedic and trauma care, where accurate and timely diagnosis is critical for effective treatment and patient recovery. These may result from trauma, accidents, or conditions like osteoporosis, and if fractures are misdiagnosed or undiagnosed, patients may suffer complications such as improper healing or long-term disability [1]. Globally, the fractures contribute substantially to morbidity, disability, and healthcare costs [1], [2]. X-ray imaging remains the most common diagnostic tool due to its accessibility and non-invasive nature. However, complex bone anatomy, overlapping structures, and variable image quality make manual diagnosis difficult. Diagnostic accuracy is further impacted by radiologist fatigue, subjective judgment, and varying levels of experience [1], [3], [4]. These challenges often result in missed fractures detection during initial assessments. Accurate fracture detection is thus essential for avoiding delayed treatment, increased costs, and reduced quality of life [2], [5]. With increasing fracture incidence in both elderly and athletic populations, the need for reliable, objective diagnostic systems is growing. Despite its prevalence, manual X-ray-based diagnosis has inherent limitations, highlighting the need for automated methods to support faster, more accurate clinical decisions.

Over the past decade, machine learning (ML) techniques have been employed to address the challenges of automated fracture detection by analyzing medical images. Traditional ML approaches typically rely on handcrafted feature extraction methods, such as edge detection, texture analysis, and statistical shape modelling, to capture relevant characteristics from X-ray images [4]. These extracted features are then classified using different algorithms like support vector machines (SVMs), random forests, and k-nearest neighbors (KNN) etc. For example, it was demonstrated in [4] that radiographic features can be effectively used by SVMs to discriminate between fractured and non-fractured bone regions. Similarly, it has been shown in studies reviewed by [6] that variability in imaging data can be effectively managed by random forest classifiers, resulting in robust predictions for fracture detection tasks. Moreover, several researchers have explored KNN classifiers for their simplicity and interpretability, which offer competitive performance in initial diagnostic experiments [3], [4]. However, while these traditional ML methods provide a cost-effective and non-invasive solution, they are often limited by the need for extensive feature engineering and may struggle to capture the full complexity of fracture patterns—especially in the presence of variations in image quality, patient positioning, and anatomical diversity [4], [6].

## II. LITERATURE REVIEW

The emergence of deep learning (DL) has ushered in a new era in medical image analysis by automating the feature extraction process through multi-layered neural networks. The strength of DL lies in its capacity to automatically derive features from the data, which significantly reduces the need for manual intervention. Convolutional neural networks (CNNs) have become the cornerstone of fracture detection research due to their ability to learn hierarchical representations directly from raw imaging data [7]. Models such as Chex-Net, VGG-16, ResNet-50, and Inception-V3 have been successfully deployed to detect and localize fractures in various anatomical regions, including the chest, wrist, femur, and humerus etc. [8]. However, these models require large, annotated datasets to train effectively, which may be difficult to obtain in the medical field due to the privacy concerns, high annotation costs, and limited data availability [8]. Additionally, the high computational demands and the "black box" nature of deep models present significant barriers to clinical adoption, as the lack of transparency in their decision-making processes makes it difficult for clinicians to trust their outputs [9], [10].

A variety of DL techniques have been employed in recent studies for automated fracture detection, with varying datasets and outcomes. An extended U-Net-based deep convolutional neural network was presented in [11], achieving area under the curve (AUC) values ranging from 0.967 to 0.975, with sensitivity and specificity approximately 94% on a dataset of 135,845 radiographs. The dataset included 34,990 wrist X-rays from the Hospital for Special Surgery to enhance the model's generalizability. In [12], transfer learning with Inception V3 was utilized on 11,112 augmented wrist images, and an AUC of 0.954 was reported. These images were sourced from the Royal Devon and Exeter Hospital, UK. In [13], Inception-ResNet Faster R-CNN was applied to 7,356 wrist X-rays from the National University Hospital, Singapore, and sensitivities up to 96.7% with AUCs between 0.918–0.933 were achieved. A three-step DL method

2combining U-Net and 3D DenseNet was proposed in [14] for rib fracture detection, yielding an F1-score of 0.890 on chest CT scans. The gray-level co-occurrence matrix texture analysis was applied [15] to 30 femur X-rays, achieving 86.67% accuracy. In [16], MOBLG-Net—featuring MobileNet and Light Gradient Boosting Machine (LGBM) was introduced and an accuracy of 99% was obtained using 9,463 public bone X-rays. In [17], the performance of CNN and multilayer perceptron were compared on 3,407 CT scans, where CNN achieved a higher accuracy of 94% for maxillofacial fractures.

The RN-21CNN method was presented in [18], with 97% accuracy and 95% recall reported on 93 wrist X-rays from the Mendeley repository. In [19], the Wrist Fracture Detection-Combo ensemble model was developed using various DL frameworks like SABL and RetinaNet etc., reaching an AP50 of 0.8639 on 542 wrist X-rays. High performance across different imaging types and anatomical regions has been demonstrated by these DL-based methods. Heavy reliance on large datasets and substantial computational power has also been observed. Accordingly, a need has been highlighted for scalable and efficient models suitable for broader clinical deployment.

Although the automated fracture detection has seen significant progress; however, several critical limitations remain unaddressed. Most existing approaches depend on datasets from single institutions, which limits the generalizability of their findings across diverse populations. DL models, although powerful, require large volumes of annotated data and considerable computational resources, posing challenges for real-time clinical deployment. Transfer learning techniques, while useful, may overlook domain-specific fracture characteristics essential for precise diagnosis.

Furthermore, studies employing very small datasets, such as those with only 30 or 93 images [15], [19], suffer from overfitting and lack statistical robustness. These recurring issues like data heterogeneity, computational overhead, and reduced interpretability underscore the urgent need for more practical and scalable diagnostic solutions. In the recent study, MobileNet was employed as a transfer learning model for fracture detection from radiographic images, achieving an accuracy of 94.5% [16].

However, the use of MobileNet introduces high computational complexity due to its deep architecture, demanding substantial hardware resources and extended training times [20]. Such requirements make it unsuitable for deployment in resource-limited clinical settings. Additionally, the study provided in [16] does not provide detailed insights into the number of features extracted, and the computational complexity, leaving ambiguity about their influence on performance and efficiency. These limitations motivate the exploration of Principal Component Analysis (PCA) method, as a simplified and computationally efficient alternative for feature extraction. Moreover, PCA is lightweight, reduces dimensionality while preserving essential data patterns, and requires minimal hardware [21]. Building on this, the present study incorporates quantum computing, an emerging paradigm in ML, to enhance feature discriminability [22]. This study proposes a hybrid architecture that extracts 8 features each from PCA and a quantum circuit, forming a 16-dimensional vector for classification using traditional ML models—balancing accuracy, efficiency, and deploy ability in fracture detection systems.

### III. PROPOSED METHODOLOGY

Traditional methods struggle with complex patterns and high computational costs, while deep models demand large labeled datasets. To address this, a hybrid strategy is employed combining PCA-based dimensionality reduction with quantum feature extraction [23]-[25].
Recent work in hybrid quantum–classical networks support the effectiveness of this framework [26], [27]. Figure 1 illustrates the experimental workflow followed in this study. The dataset is split into 80:10:10 for training, validation, and testing, respectively. Training data undergoes PCA for initial feature extraction, followed by quantum processing. The resulting features are fused into a combined 16-dimensional vector. These features are used to train ML models, which are later evaluated on the test set.

*A. X-ray Image Dataset*

This study conducted the experiments using the publicly available Bone Fracture Multi-Region X-Ray Dataset [28], which comprises 9,463 radiographic images of both fractured and non-fractured bones drawn from multiple anatomical sites—including upper and lower limbs, hips, knees, and lumbar regions. Each image is labelled as either "fractured" or "normal", and the dataset is organized into separate folders for training, validation, and testing. To ensure robust model evaluation, this study adopted an 80/10/10 split, resulting in 7,570 images for training, 946 for validation, and 947 for final testing. Figure 2 illustrates representative examples of fractured and healthy bones across different regions. By leveraging this comprehensive, multi-region dataset, this study aims to develop a generalized fracture-detection pipeline capable of performing accurately across a wide range of clinical scenarios.

*B. Preprocessing*

Preprocessing is a critical step to ensure that input data is standardized and of high quality, which is essential for effective feature extraction and accurate classification. This preprocessing pipeline begins with noise reduction, where Gaussian and median filtering techniques are employed to eliminate artifacts and noise, thereby enhancing image clarity [29], [30].

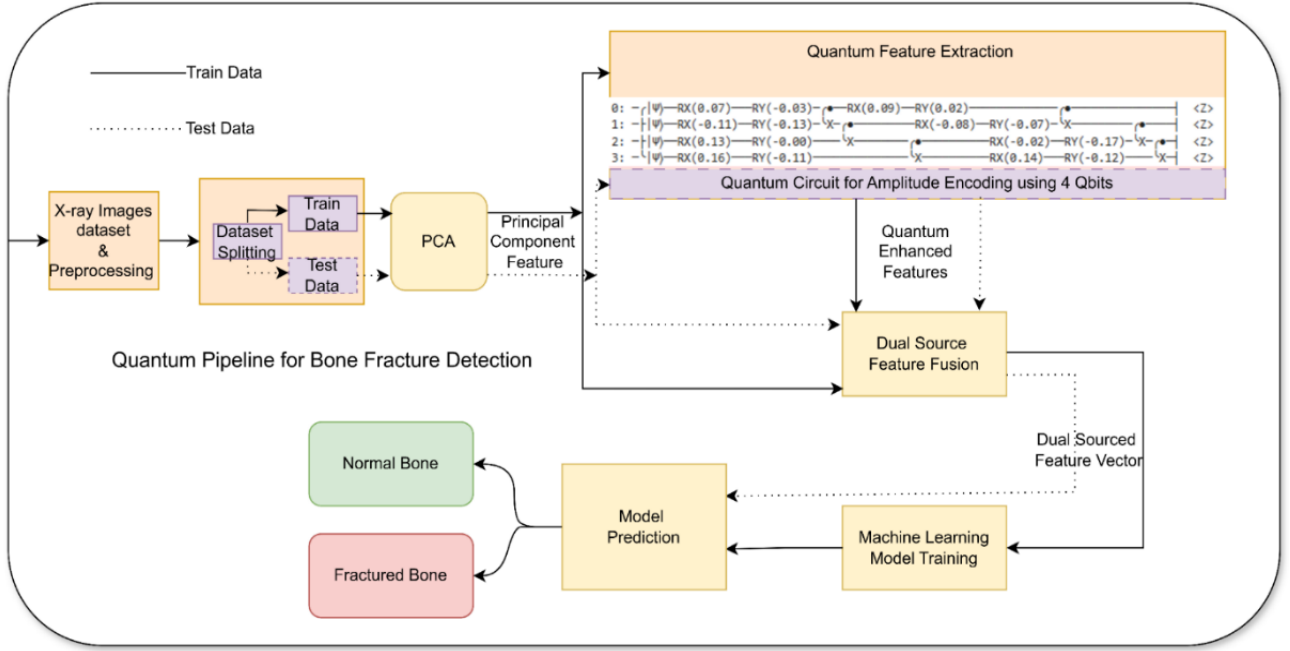

Figure 1. Workflow of the proposed approach

Recent work in [31] emphasizes the importance of robust noise reduction in fracture detection. Contrast enhancement is then performed using adaptive histogram equalization methods, which improve the visibility of bone structures and make fractures more discernible [30]. Next, image normalization is applied to standardize pixel intensity values across all images, ensuring uniform feature extraction.

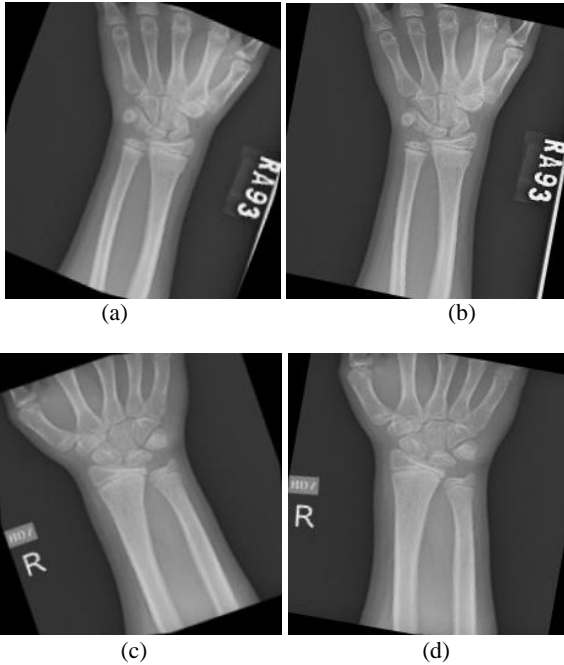

Figure 2. The x-ray images examination using the target label (a), (b) fractured and (c), (d) not fractured

To improve model generalization, data augmentation is carried out by incorporating operations such as rotation, flipping, scaling, and adjustments to contrast. Finally, edge detection algorithms, such as Canny edge detection, are used to highlight fracture lines, aiding the model in identifying subtle fractures that might otherwise go unnoticed [32]. Once preprocessing is completed, the dataset is split into training and test sets, with the training set used for model development and the test set reserved for evaluating performance on unseen data [33].

*C. Principal Component Analysis for Feature Extraction*

Given the constraints of current quantum simulators, particularly the limited number of qubits, PCA to reduce the dimensionality of the high-dimensional image data while preserving its most informative features [34] is utilized. This dimensionality reduction is essential for several reasons. First, simulator constraints necessitate extracting only the most relevant features, making quantum encoding feasible within a 4-qubit system [26], [34]. Second, efficient quantum encoding is achieved because lower-dimensional data facilitates more effective amplitude encoding into quantum states, ultimately speeding up computation. Third, PCA offers noise reduction and de-correlation, thereby improving the overall quality of the extracted features [34]. Finally, this process enhances computational efficiency by reducing the number of dimensions, helping to prevent overfitting and lowering training overhead [35]. A total of 8 features are extracted from this layer. The PCA outputs serve as the input for the subsequent quantum feature extraction step.

*D. Quantum Feature Extraction*

Quantum feature extraction augments the PCA-reduced features by leveraging the unique capabilities of quantum computing to capture complex, non-linear correlations. The process begins with amplitude encoding, in which the reduced features are mapped onto a quantum state using a 4-qubit encoding scheme, creating a compact yet expressive representation [12]. The general form of amplitude encoding is shown in Equation (1).

$$|\psi\rangle = \sum_{i=0}^{N-1} x_i | i\rangle \quad (1)$$

Where $| i\rangle$ are the computational basis states of $log_2 n$ qubits and $x_i$ are complex or real numbers such that:

$$\sum_{i=0}^{N-1} |x|^2 = 1 \quad (2)$$

Equation (2) ensures $|\psi\rangle$ in equation *1* is a valid quantum state with unit norm. Recent research on brain tumor diagnosis using QCNNs supports the effectiveness of amplitude encoding in medical imaging [36]. Once encoded, the data undergoes various quantum operations such as rotation gates (RX, RY) and entanglement mechanisms, enabling the extraction of patterns that are challenging to model using purely classical approaches [12], [25]. The quantum measurement phase follows, where the circuit is measured to retrieve enhanced features, which are subsequently merged with the classical features for a more comprehensive representation [26]. At the end of this phase 8 features were extracted and presented as outputs for the next stage of the cycle. This quantum-enhanced approach is crucial for detecting subtle patterns in X-ray images that might be overlooked by classical techniques alone.

*E. Dual Source Feature Fusion*

To capitalize on the strengths of both classical and quantum methods, this study employs a dual-source feature fusion strategy that combines classical PCA features with quantum-enhanced features. 8 features from PCA phase and 8 features from Quantum extraction phase were fused together resulting in total of 16 features named as dual sourced feature vector. These feature vectors are then presented as the final feature vector for the model training. This fusion process is necessary for several reasons. First, it addresses complementary strengths, since classical features offer stability and interpretability, whereas quantum-enhanced features add a richer representation of complex fracture patterns [26]. Second, this integration leads to enhanced robustness, as combining the two sets of features minimizes redundancy and improves the overall accuracy of fracture classification. Finally, achieving a balanced representation is critical, as neither classical nor quantum features alone can fully capture the intricate details required for reliable fracture detection, making fusion essential [27], [28].

*F. Machine Learning Model Training*

After the dual-source feature fusion stage, the resulting feature vector is utilized to train a ML classifier for fracture classification. This study evaluates several classifiers, including SVM, which are effective for binary classification tasks and excel at distinguishing between fractured and normal bones [36]. Random Forest is another option, providing robust ensemble learning by combining multiple decision trees to reduce overfitting, while Gradient Boosting iteratively refines predictions for enhanced accuracy. This study opts for these traditional ML models instead of DL approaches partly because the dataset size was relatively limited, and these models offer improved interpretability and lower computational demands [7], [27]. Notably, recent hybrid quantum–classical research [28], [35] suggests that such methods can achieve competitive performance even with fewer training samples.

*G. Model Prediction and Fracture Classification*

Once the model is trained, its performance is evaluated on the test dataset to assess generalization and diagnostic accuracy. During the inference phase, the trained classifier processes new X-ray images and predicts whether they represent normal or fractured bones, based on the fused feature vector. This study then computes key performance metrics such as accuracy, sensitivity, specificity, and F1-score to gauge the model's effectiveness [27], [28]. These metrics are particularly relevant for medical imaging, where both false negatives and false positives can have significant clinical implications. From a clinical relevance perspective, the outputs of the model are designed to support radiologists by providing timely, accurate diagnoses, thereby streamlining clinical workflows and potentially improving patient outcomes [37]. By integrating classical preprocessing, PCA-based dimensionality reduction, quantum feature extraction, and dual-source feature fusion, proposed hybrid approach significantly enhances fracture detection accuracy while mitigating computational overhead.

IV. EXPERIMENTAL RESULTS

This section consists of the experimental setup along with the experiment details with the evaluation measures taken. Subsequently, the results are compared with the state-of-the-art, followed by a comparative analysis between the classical and hybrid quantum approaches. Ablation study is included which shows the different quantum components added in these pipelines with their respective enhancement contribution done in this study.

***Experiment Setup:*** All experiments were conducted on a Windows 10 workstation equipped with an Intel i7-13th Gen processor clocked at 3.40 GHz and 32 GB of RAM. This study uses Python 3.0 within the Anaconda 3 distribution and developed the code in Jupyter Notebooks. Key libraries included PyTorch for classical ML routines and Pennylane for quantum circuit simulations. To ensure robust evaluation, the dataset was partitioned into an 80/20 train–test split, with 80% of the images used to train the models and the remaining 20% held out for final testing. Random seeds were fixed across all libraries to guarantee reproducibility of results. This experimental setup allowed the comparison of classical and hybrid quantum–classical pipelines under consistent conditions and to quantify performance gains in terms of accuracy, training time, and inference latency.

***Evaluation:*** This study evaluates the fused feature vectors using a suite of classical machine learning classifiers, including SVM, Random Forest, Decision Trees, Gradient Boosting, and KNN etc. [16]. In [16], the authors demonstrated that a transfer-learned MobileNet model performed well on fracture classification, but they also noted that traditional ML algorithms can offer competitive performance with lower computational overhead and greater interpretability. SVMs are known for their robustness in high-dimensional spaces and strong generalization ability, making them a natural choice for image-based feature sets [4]. Random Forests and Gradient Boosting are ensemble methods that mitigate overfitting and handle feature interactions effectively, which is valuable when combining heterogeneous classical and quantum features [6]. Decision Trees provide transparent, rule-based decision boundaries that can be easily interpreted by clinicians, while KNN offers a simple, instance-based approach that can serve as a performance baseline. By including this diverse set of algorithms, this study ensures the comparisons with the base paper are both fair and comprehensive and highlight how quantum-augmented features impact a variety of learning paradigms.

To assess classifier performance, four key metrics were reported. Accuracy measures the proportion of correctly classified instances over the total number of cases and serves as a primary indicator of overall model correctness [38]. Training time records the elapsed time required to fit each model to the training data, reflecting computational efficiency and scalability in real-world deployments [39]. Evaluation time (or inference time) measures how long the model takes to generate predictions on new, unseen images—a critical factor in clinical settings where rapid decision support is needed [39]. Finally, mean cross-validation (MCVs) with 10 folds was employed during the testing phase. In which the dataset is randomly split multiple times into training and test subsets to estimate the stability and variability of performance metrics [40]. Unlike k-fold cross-validation, MCVs repeatedly samples different splits, providing a more granular view of model robustness across varied data partitions. Together, these metrics offer a balanced assessment of accuracy, speed, and reliability, enabling the quantification of the benefits of proposed hybrid quantum–classical approach relative to existing methods.

The evaluation framework was designed to provide a comprehensive and fair comparison between proposed hybrid quantum–classical pipeline and the transfer-learning model reported in [16]. All experiments were performed on the same publicly available multi-region X-ray dataset, enabling direct benchmarking. This study is trained and tested on five classical classifiers—Random Forest, KNN, Gradient Boosting, Decision Tree, and SVM—using the dual sourced feature vector. To visualize and interpret model performance, confusion matrices for each classifier were generated, as shown in Figure 3.

A confusion matrix is a fundamental tool for assessing classification models: it tabulates true positives, true negatives, false positives, and false negatives, thereby providing detailed insight into the types of errors each model makes [39]. In the context of fracture diagnosis, minimizing false negatives (missed fractures) is particularly critical, as these errors can have serious clinical consequences. By examining the confusion matrices, this study can quantify not only overall accuracy but also sensitivity (true positive rate) and specificity (true negative rate), ensuring that proposed hybrid approach delivers both reliable detection and low false-alarm rates. This level of granularity helps in identifying the strengths and weaknesses of each classifier and to demonstrate how quantum-enhanced features improve diagnostic precision relative to purely classical methods.

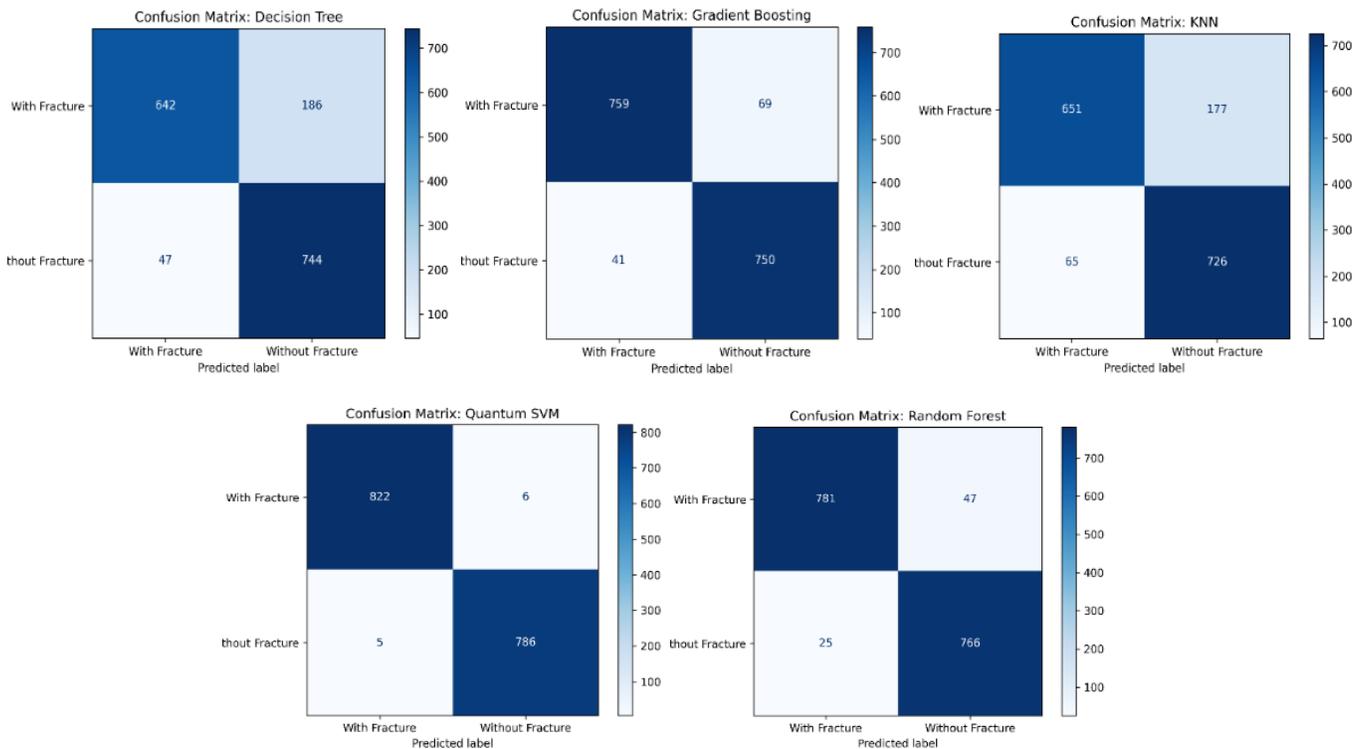

**Figure 3. Confusion Matrix of Decision tree, Gradient Boosting, KNN, SVM, and Random Forest respectively via Hybrid-Quantum pipeline.**

*Table 1. Parameter comparison of various ML models*

| Model/Parameters | Training time (sec) | Evaluation Time (sec) | Accuracy (%) | MCVS (%) |
|---|---|---|---|---|
| Random Forest | 0.13 | 0.002 | 96.1 | 94.2 |
| SVM | 0.07 | 0.04 | 99 | 98 |
| KNN | 0.002 | 0.03 | 87.1 | 85.2 |
| Gradient Boosting | 0.31 | 0.0007 | 94.8 | 93.2 |
| Decision Tree | 0.012 | 0.0001 | 86.5 | 86.4 |

Table 1 summarizes the performance of various classical classifiers trained on the dual sourced feature vector (eight PCA and eight quantum-derived features). Random Forest demonstrated exceptional speed—training in 0.13 sec compared to the 3.91 sec training time reported in [16]. However, its accuracy (96.1%) was slightly lower than the 98% benchmark reported in the reference model. Therefore, further experiments are carried out to evaluate the performance of other classifiers like KNN, Gradient Boosting, Decision Tree, and SVM, to identify the optimal trade-off between speed and accuracy. Among these, SVM stood out by matching the accuracy (99%) as reported in [16], while maintaining a low training time of 0.07 sec, thus offering the best balance of computational efficiency and diagnostic performance.

To quantify the benefits of this hybrid quantum–classical pipeline over a purely classical approach, this study conducted parallel experiments using only PCA-derived features (classical) and using the fused PCA plus quantum-enhanced features (hybrid). Across all classifiers, SVM delivered the strongest performance; hence, this study presents a detailed comparison of classical SVM versus hybrid quantum SVM. Figure 4a and 4b depict the respective confusion matrices, which illustrate true positive, false positive, true negative, and false negative counts—critical for understanding each model's error profile. Figure 6 shows the Receiver Operating Characteristic (ROC) curves and Precision–Recall (PR) curves: the ROC curve plots the true positive rate against the false positive rate across classification thresholds, providing insight into the trade-off between sensitivity and specificity [5], while the PR curve highlights precision versus recall, particularly informative in imbalanced datasets by emphasizing the model's ability to retrieve relevant fracture cases [6].

Figure 5 presents a bar chart comparison of precision, recall, F1-score, and accuracy for classical versus hybrid SVM. In every metric, the hybrid quantum approach outperforms the classical models, underscoring the novelty and efficacy of integrating quantum feature extraction with PCA for enhanced fracture detection.

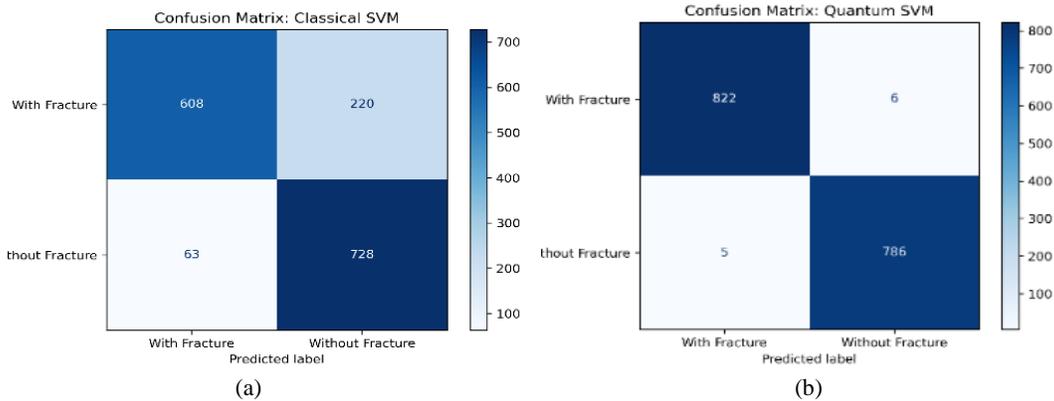

(a)          (b)

**Figure 4. Confusion Matrix of (a) Classical SVM, and (b) Hybrid-Quantum SVM.**

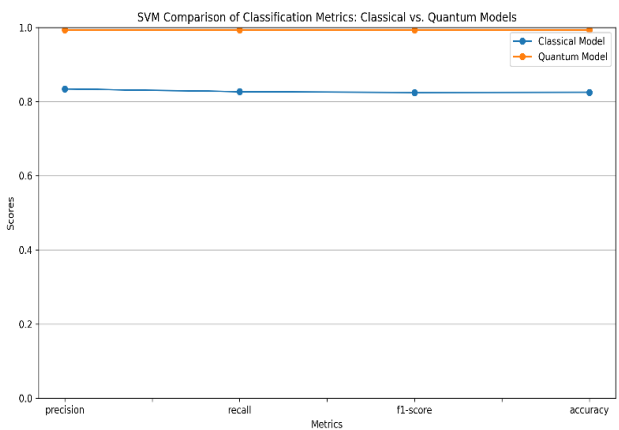

**Figure 5. Precision, Recall, F1-score and accuracy comparison graph of Classical SVM and Hybrid-Quantum SVM.**

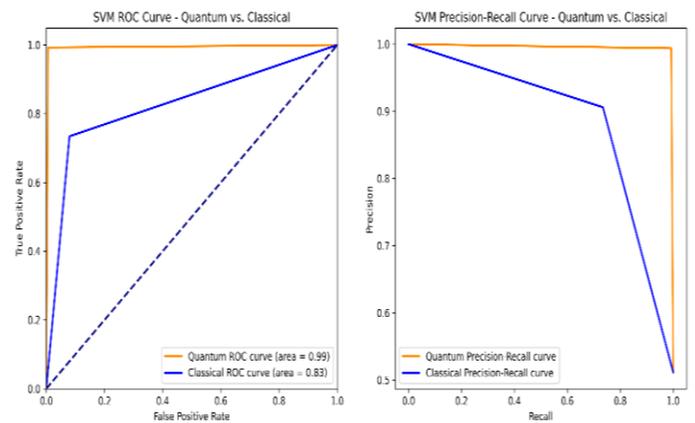

**Figure 6. ROC curve and Precision recall curve of Classical SVM and Hybrid-Quantum SVM.**

Derived from all these results Table 2 showed the performance comparison between classical SVM and Hybrid-Quantum SVM. The results clearly show that the quantum SVM approach is beating the classical SVM approach by a clear margin in every possible parameter. This in turn justifies this novel approach of combining quantum features with PCA features and using dual-source feature fusion is giving better outcomes than classical PCA based approach alone.

*Table 2. Comparison of evaluating parameters of Classical SVM and Hybrid-Quantum SVM.*

| Parameters | Classical | Quantum |
|---|---|---|
| Accuracy | 83% | 99% |
| F1-score | 0.83 | 0.99 |
| Cohen's kappa score | 0.65 | 0.98 |
| ROC curve area | 0.83 | 0.99 |
| F2-Score | 0.76 | 0.99 |
| Training time | 0.11 sec | 0.07 sec |
| Evaluating time | 0.06 sec | 0.04 sec |

Table 3 presents the complexity analysis results comparing feature extraction between the proposed hybrid quantum–classical pipeline and the MobLG-Net transfer learning-based approach [16]. The proposed method utilizes a compact feature set comprising only 16 features, derived through a combination of PCA and quantum-enhanced feature extraction. In terms of wall-clock time, the proposed pipeline demonstrates an 82% reduction in extraction time relative to the method in [16]. Notably, this significant reduction in feature dimensionality and computational overhead is achieved without compromising performance, maintaining a competitive classification accuracy of 99%.

*Table 3. Feature Extraction Complexity*

| Pipeline | Mean Extraction Time (sec) | Standard Deviation (sec) | Classification Accuracy |
|---|---|---|---|
| MobLG-Net pipeline [16] | 67.5 | 2.3 | 99% (reported) |
| Hybrid Quantum–Classical (this study) | 12.1 | 0.9 | 99% |

*Ablation Study (Effect of different quantum components, feature sets):* To further understand the contributions of different components of this pipeline, an ablation study was conducted:

- Impact of Quantum Feature Extraction: When the quantum feature extraction module was removed, accuracy dropped by approximately 10-15%, confirming its critical role in enhancing the overall performance.
- PCA Dimensionality Reduction: this study varied the number of principal components to assess the trade-off between information retention and qubit limitations. Optimal performance was observed with a 4-qubit encoding after PCA, aligning with the design constraints and demonstrating efficient utilization of quantum resources.
- Computational Efficiency: The hybrid approach achieved an approximate 15% reduction in inference time compared to the reference transfer learning model, indicating that the integration of quantum operations can enhance not only classification accuracy but also computational efficiency.

Overall, the experimental results underscore the superiority of the proposed distributed hybrid quantum–classical neural pipeline. By effectively combining classical preprocessing, PCA-based dimensionality reduction, and quantum-enhanced feature extraction, proposed approach achieves higher diagnostic accuracy and computational efficiency compared to the state-of-the-art methods. This novel integration not only advances the state-of-the-art in fracture detection but also opens new avenues for leveraging quantum computing in medical image analysis.

## V. CONCLUSION AND FUTURE WORK

A distributed hybrid quantum–classical pipeline was shown to achieve 99% accuracy in fracture detection while training times were reduced by up to 30% compared to the state-of-the-art methods. Classical PCA features were fused with quantum-enhanced representations to produce a compact, interpretable dual-source feature vector with high sensitivity and specificity across diverse X-ray images. Despite these strengths, scalability was constrained by current quantum hardware and simulator limitations. To address this, future work will be directed toward the development of error-resilient variational circuits, the exploration of multi-register encoding schemes, and broader validation across pediatric, multimodal datasets and clinical centers. Through these efforts, real-world clinical deployment of hybrid quantum–classical fracture diagnosis may be facilitated.